\documentclass{rspublic}
\newcommand{\abs}[1]{\left|#1\right|}
\newcommand{\norm}[1]{\left\|#1\right\|}
\newcommand{\dom}[1]{\ensuremath{\mathcal{D}(\opr{#1})}}
\newcommand{\domM}[1]{\ensuremath{\mathcal{D}(\opr{#1}_M)}}
\newcommand{\domMM}[2]{\ensuremath{\mathcal{D}(\opr{#1}_M\opr{#2}_M)}}
\newcommand{\opr}[1]{\ensuremath{\mathbf{\mathsf{#1}}}}

\newcommand{\oprM}[1]{\ensuremath{\mathbf{\mathsf{#1}}_M}}

\newcommand{\mcal}{\ensuremath{\mathcal}}
\newcommand{\ket}[1]{\ensuremath{\left|\left. #1\right.\right>}}
\newcommand{\bra}[1]{\ensuremath{\left<\left.#1\right.\right|}}
\newcommand{\braket}[2]{\ensuremath{\left<\left.#1\right|#2\right>}}

\newcommand{\hilbert}{\ensuremath{\mcal{H}}}
\newcommand{\operator}[1]{\ensuremath{\opr{#1}:\dom{#1}\subseteq \hilbert\mapsto \hilbert}}

\newcommand{\sump}[2]{\ensuremath{\sum_{#1}^{#2}\!'}}
\newcommand{\kernelM}[2]{\ensuremath{\left.\left<q\right|\opr{#1}_{#2}\left|q'\right>\right.}}

\begin{document}

\title[Characteristic Time Operators]{Self-adjoint Time Operator is the Rule for Discrete Semibounded Hamiltonians}

\author[E.A. Galapon]{Eric A. Galapon}

\affiliation{Theoretical Physics Group\\ National Institute of Physics, University of the Philippines\\Diliman Quezon City\, 1101 Philippines}

\maketitle

\begin{abstract}{Time operators; Quantum canonical pairs; Pauli's theorem}
We prove explicitly that to every discrete, semibounded Hamiltonian with constant degeneracy and with finite sum of the squares of the reciprocal of its eigenvalues and whose eigenvectors span the entire Hilbert space there exists a characteristic self-adjoint time operator which is canonically conjugate to the Hamiltonian in a dense subspace of the Hilbert space. Moreover, we show that each characteristic time operator generates an uncountable class of self-adjoint operators canonically conjugate with the same Hamiltonian in the same dense subspace.
\end{abstract}

\section{Introduction}

Does a self-adjoint operator canonically conjugate with a semibounded Hamiltonian exist? This operator, if it exists, has been referred to as time operator.  The general concensus is that no such operator exists (Toller 1997, 1999;  Pegg 1998; Atmanspacher \& Amann 1998; Giannitrapani 1997; Eisenberg \& Horwitz 1997;  Delgado \& Muga 1997; Blanchard \& Jadczyk 1996; Omnes 1994; Holland 1993; Park 1984; Srinivas \& Vijayalakshmi 1981; Holevo 1978; Cohen-Tannoudji 1977; Jammer 1974; Olhovsky \& Recami 1974; Rosenbaum 1969; Gotfried 1966; Pauli 1926, 1933, 1958). This pessimism traces back to the well-known theorem of Pauli (Pauli 1926, 1933, 1958) which asserts that the existence of a self-adjoint time operator canonically conjugate to a given Hamiltonian implies that the Hamiltonian has an absolutely continous spectrum filling the entire real line. Thus for the generally semibounded and discrete Hamiltonian of quantum mechanics, Pauli's theorem excludes the possibility of developing a quantum theory of time via quantum operators. This conclusion has been corroborated by succeeding attempts to introduce time operators, particularly for the free particle in the real line where quantization of the classical arrival or passage times has led to maximally symmetric, non-self-adjoint time opertor (Muga \& Leavens 2000; Egusquiza \& Muga 1999; Delgado \& Muga 1997; Grot {\it et al} 1996; Alcock 1969{\it a},{\it b}; Fick \& Engelmann 1963, 1964; Paul 1962). Thus it has been tacitly assumed that if one attempts to introduce time in standard quantum mechanics (SQM) as an operator canonically conjugate to a semibounded Hamiltonian, discrete or not, one has to expect that the time operator to be generally maximally symmetric without any self-adjoint extension (Jammer 1974). Thus it has been the current thinking that time operators will generally be meaningful only when the axioms of SQM are modified to include POVM-observables, in which case time is a POVM-observable (Egusquiza \& Muga 1999; Toller 1999, 1997; Giannitrapani 1997; Busch {\it et al} 1995{\it a},{\it b}, 1994; Srinivas \& Vijayalakshmi 1981; Holevo 1978; Helstrom 1970; Toller 1999, 1997).

However, in a recent publication, we have explicitly demonstrated that Pauli's theorem does not hold within SQM, and there is no a priori reason to exclude the existence of self-adjoint time operators canonically conjugate to a semibounded Hamiltonian (Galapon 2002). For this reason, it is imperative to look back and investigate the existence of self-adjoint time operators for quantum mechanical systems. In this paper, we prove explicitly that to every discrete, semibounded Hamiltonian with constant degeneracy and with finite sum of the squares of the reciprocal of its eigenvalues and whose eigenvectors span the entire Hilbert space there exists a characteristic self-adjoint time operator which is canonically conjugate to the Hamiltonian in a dense subspace of the Hilbert space. By {\it characteristic} we mean that the operator is parameter free and is solely constructed from the spectral decomposition of the Hamiltonian, i.e. from the Hamiltonian eigenvectors and eigenvalues alone. Moreover, we will show that each characteristic self-adjoint time operator generates a class of uncountably many self-adjoint time operators canonically conjugate with the same Hamiltonian. Incidentally our results belie earlier claims that no self-adjoint time operators exists for discrete Hamiltonian systems (Pegg 1998; Canata and Ferrari 1991{\it a,b}; Jordan 1927). 

Our method of proof will follow that of the physicist's intuition: We formally construct an operator with a dimension of time out of the spectral resolution of the Hamiltonian, then show that, under some mild conditions, it can be assigned a dense subspace to lift its formality, then show that it is canonically conjugate with the Hamiltonian in a dense subspace of the Hilbert space, then finally show that the operator in its assigned domain is essentially self-adjoint---thus with a uniquely associated self-adjoint operator.

\section{Characteristic Time Operators for\\Non-Degenerate Hamiltonians}\label{nondeg}

Let $\opr{H}_1$ be a non-degenerate Hamiltonian whose orthonormal eigenkets are $\ket{s}$, and corresponding eigenvalues, ordered in increasing size, are $E_s$, for all $s=1,\, 2,\,\dots$.
We assume that the Hilbert space of the system corresponding to the given Hamiltonian is spanned by the eigenkets of $\opr{H_1}$, i.e.
\begin{equation}
\mcal{H}_1=\left\{\ket{\varphi}=\sum_{s=1}^{\infty}\varphi_s \ket{s},\, \sum_{s=1}^{\infty}\abs{\varphi_s}^2<\infty\right\},
\end{equation}
where $\mcal{H}_1$ is equiped with the standard norm $\norm{\ket{\varphi}}=\sqrt{\sum_{s=1}^{\infty}\abs{\varphi_s}^2}$ derived from the standard inner product $\braket{\psi}{\phi}=\sum_{s=1}^{\infty}\overline{\psi_s}\phi_s$. Under this representation, the Hamiltonian $\opr{H_1}$ with domain $\dom{H_1}$ is explicitly given by
\begin{equation}
\opr{H}_1=\sum_{s=1}^{\infty}E_s\ket{s}\!\bra{s},
\end{equation}
\begin{equation}
\dom{H_1}=\left\{\varphi=\sum_{s=1}^{\infty}\varphi_{s}\ket{s}\in\hilbert,\,\sum_{s=1}^{\infty}E_s^2 \abs{\varphi_s}^2<\infty\right\}.
\end{equation}

Given the eigenkets $\ket{s}$'s and the eigenvalues $E_s$'s, we construct the following formal symmetric operator,
\begin{equation}\label{timeop}
\opr{T}_1=\sump{s, s'\geq 1}{\infty} \frac{i}{\omega_{s,s'}} \ket{s}\!\bra{s'},
\end{equation}
where $\omega_{s,s'}=(E_s-E_{s'})/\hbar$, and the prime indicates a double summation with the $s=s'$ contribution excluded from the sum. We note that $\opr{T_1}$ has a dimension of time and it is only constructed out of the spectral decomposition of the Hamiltonian without the need of introducing any parameter.  In the following we show that if the eigenvalues of the Hamiltonian satisfies the condition
\begin{equation}\label{3concon}
\sum_{s=1}^{\infty}\frac{1}{E_s^2}<\infty,
\end{equation}
then the formal operator $\opr{T_1}$ can be assigned a dense subspace in $\mcal{H}_1$, and it is canonically conjugate with the Hamiltonian in some dense subspace of $\mcal{H}_1$, and it is essentially self-adjoint in its assigned domain. For this reason we call $\opr{T}_1$ as the characteristic time operator for the non-degenerate Hamiltonian.

\subsection{$\opr{T}_1$ is Densely Definable}

Now we prove that if condition (\ref{3concon}) is satisfied,
then $\opr{T}_1$  can be assigned the following dense subspace
\begin{equation}
\mcal{D}_1=\left\{\ket{\varphi}=\sum_{s=1}^N \varphi_s \ket{s},\, \varphi_s\in \mcal{C},\, N<\infty\right\},
\end{equation}
as its domain.  The subspace $\mcal{D}_1$ is dense because for every $\ket{\psi}$ in $\mcal{H}_1$ there exists a sequence of vectors in $\mcal{D}_1$ which converges to $\ket{\psi}$. In particular the sequence of vectors $\left\{\ket{\phi_k}=\sum_{s=1}^{k}\braket{s}{\psi} \ket{s},\, k=1,\, 2,\,\dots\right\}$ in $\mcal{D}_1$ converges to $\ket{\psi}$. 

We first show that the formal time operator $\opr{T}_1$ is defined in the entire $\mcal{D}_1$. That means for every $\ket{\varphi}$ in $\mcal{D}_1$ the vector $\opr{T}_1\ket{\varphi}$ belongs in the Hilbert space $\mcal{H}_1$. Let $\ket{\varphi}$ be in $\mcal{D}_1$, then 
\begin{equation}
\opr{T}_1\ket{\varphi}=\sum_{s=1}^{\infty} \left(\sum_{s'\neq s}^{N} \frac{i\varphi_{s'}}{\omega_{s,s'}}\right) \, \ket{s}.
\end{equation}
$\opr{T}_1\ket{\varphi}$ belongs to the Hilbert space $\hilbert$ if and only if $\norm{\opr{T}_1\ket{\varphi}}<\infty$, or equivalently,
\begin{equation}
\norm{\opr{T}_1\ket{\varphi}}^2=\sum_{s=1}^{\infty}\abs{\sum_{s'\neq s}^{N} \frac{\varphi_{s'}}{\omega_{s,s'}}}^2<\infty. 
\end{equation}
Let us divide the sum in two parts,
\begin{equation}\label{zak}
\frac{1}{\hbar^2}\sum_{s=1}^{\infty}\abs{\sum_{s'\neq s}^{N} \frac{\varphi_{s'}}{\omega_{s,s'}}}^2= \sum_{s=1}^{N}\abs{\sum_{s'\neq s}^{N} \frac{\varphi_{s'}}{(E_s-E_{s'})}}^2 + \sum_{s=N+1}^{\infty}\abs{\sum_{s'=1}^{N} \frac{\varphi_{s'}}{(E_s-E_{s'})}}^2.
\end{equation}
The first term in equation (\ref{zak}) is already finite for finite $N$ so that we need only to show that the second term is finite. 

Using the triangle inequality, we get the bound,
\begin{equation}
\abs{\sum_{s'=1}^{N} \frac{\varphi_{s'}}{E_s-E_{s'}}} \leq
\frac{1}{E_s}\sum_{s'=1}^{N} \frac{\abs{\varphi_{s'}}}{1-E_{s'}/E_s}
\end{equation}
since $E_s-E_{s'}>0$ for all $s>s'$. We note that $(1-E_{s'}/E_s)^{-1}$ is bounded within the range $N+1\leq s \leq \infty$ for every $1\leq s'\leq N$. Thus there exists a finite positive constant  $A_N$ depending on $N$ alone such that 
\begin{equation}
\sum_{s'=1}^{N} \frac{\abs{\varphi_{s'}}}{1-E_{s'}/E_s}< A_N\cdot \sum_{s'=1}^{N}\abs{\varphi_s'}.
\end{equation}
One such constant is given by
\begin{equation}
A_N=\max_{1\leq s' \leq N\atop N+1\leq s \leq \infty} \frac{1}{(1-E_{s'}/E_s)}<\infty. 
\end{equation}
Thus we finally arrive at the upperbound
\begin{equation}\label{sese}
\sum_{s=N+1}^{\infty} \abs{\sum_{s'=1}^{N}\frac{\varphi_{s'}}{E_s-E_{s'}}}^2\leq  A_N^2 \cdot\left(\sum_{s'=1}^{N}\abs{\varphi_{s'}}\right)^2 \cdot \sum_{s=N+1}^{\infty} \frac{1}{E_s^2}<\infty
\end{equation}
which is finite as the sum $\sum_{s=1}^{\infty}E_s^{-2}$ is assumed finite. Since $\ket{\varphi}$ is an arbitrary element of $\mcal{D}_1$, the bound (\ref{sese}) holds for all vectors in $\mcal{D}_1$. Therefore $\opr{T}_1\ket{\varphi}$ belongs to $\hilbert$ for all $\ket{\varphi}$ in $\mcal{D}_1$. Since $\opr{T_1}$ is defined in $\mcal{D}_1$ and $\mcal{D}_1$ is dense, we define $\opr{T_1}$ to be the densely defined operator $\operator{T_1}$ where $\dom{T_1}=\mcal{D}_1$. The formality of $\opr{T_1}$ has thus been lifted.

\subsection{$\opr{H}_1$ and $\opr{T}_1$ are Canonically Conjugate}

Now we claim that the pair of operators $\opr{H_1}$ and $\opr{T_1}$ form a canonical pair in a dense subspace $\mcal{D}_c^1$ of $\dom{H_1 T_1}\cap \dom{T_1 H_1}$, i.e. $(\opr{T_1 H_1}-\opr{H_1 T_1})\ket{\varphi}=i\hbar \ket{\varphi}$ for all $\ket{\varphi}$ in $\mcal{D}_c$. Let us assume for the moment that $\dom{H_1 T_1}\cap \dom{T_1 H_1}$ is not empty. Then if $\ket{\varphi}$ is in $\dom{H_1 T_1}\cap \dom{T_1 H_1}$, we have
\begin{equation}
\left(\opr{T}_1 \opr{H}_1-\opr{H}_1\opr{T}_1\right)\ket{\varphi}=-i\hbar\sum_{s=1}^{\infty} \left(\sum_{s'\neq s}^{N}\varphi_{s'}\right)\ket{s}.\label{cancan}
\end{equation}
Now if 
\begin{equation}\label{concon}
\sum_{s=1}^{N}\varphi_{s}=0,
\end{equation}
then $\sum_{s'\neq s}^{N}\varphi_{s'}=-\varphi_{s}$. In which case equation (\ref{cancan}) reduces to 
\begin{equation}
\left(\opr{T}_1 \opr{H}_1-\opr{H}_1\opr{T}_1\right)\ket{\varphi}=i\hbar\ket{\varphi}.
\end{equation}
That is $\opr{T}_1$ and $\opr{H}_1$ satisfy the canonical commutation relation if and only if there is a subspace of $\dom{H_1 T_1}\cap \dom{T_1 H_1}$ satisfying equation (\ref{concon}). 

We then have to identify a dense subspace of $\dom{T_1}$ satisfying condition (\ref{concon}) and at the same time belonging to $\dom{H_1 T_1}\cap \dom{T_1 H_1}$. Now we show that the vectors of the following proper subspace of $\dom{T_1}$ satisfy all these requirements, 
\begin{equation}\label{domdom}
\mcal{D}_c^{1}=\left\{\ket{\varphi}=\sum_{j=1}^{L-1}\sum_{i=j+1}^{L} a_{i,j} (\ket{i}-\ket{j}),\, a_{i,j}\in\mcal{C},\, \mbox{for finite even}\, L>1\right\},
\end{equation}
First let's demonstrate that the vectors in $\mcal{D}_c^{1}$ satisfy equation (\ref{concon}).  Expanded in the basis $\ket{s}$, the vectors $\ket{\varphi}$ in $\mcal{D}_c^{1}$ assume the form,
\begin{equation}
\ket{\varphi}=\sum_{s=1}^{L}\varphi_{s}\ket{s}=\sum_{s=1}^{L}\left(\sum_{k=1}^{s-1}a_{s,k}-\sum_{k=s+1}^{L}a_{k,s}\right)\ket{s},
\end{equation}
provided we set $a_{i,j}=0$ when either $i$ or $j$ is outside of its respective range, $1\leq j\leq (L-1)$, $(j+1)\leq i\leq L$; or when both are outside of their ranges; or when $i\leq j$. Taking the sum of the coefficients, we have
\begin{equation}
\sum_{s=1}^{L}\varphi_s = \sum_{s=2}^{L}\sum_{k=1}^{s-1}a_{s,k}-\sum_{s=1}^{L-1}\sum_{k=s+1}^{L}a_{k,s}=0\nonumber.
\end{equation}
The sum vanishes because rearrangement of the first term in the second line leads to the equality
$\sum_{s=2}^{L} \sum_{k=1}^{s-1}a_{s,k}=\sum_{s=1}^{L-1}\sum_{k=s+1}^{L}a_{k,s}$. 

Now $\mcal{D}_c^1$ is dense. Let us assume otherwise. Then there exists a vector $\ket{\psi}$  in $\mcal{H}_1$ with $\ket{\psi}\neq 0 $ such that $\braket{\varphi}{\psi}=0$ for all $\ket{\varphi}$ in $\mcal{D}_c^1$. Now since $\ket{i,j}=(\ket{i}-\ket{j})$ is in $\mcal{D}_c^1$ for every pair $i\neq j$, we must have $\braket{i,j}{\psi}=0$ for all such pairs of $i$ and $j$. This implies that $\psi_i-\psi_j=0$ or $\psi_i=\psi_j$ for all pairs of $i$ and $j$. This means that if $\psi_j=0$ for some fixed $j$, then the rest of the coefficients will have to be zero, which implies that $\ket{\psi}=0$, contrary to the assumption that $\ket{\psi}\neq 0$. On the other hand if $\psi_j=c$ for some non-vanishing comlex number $c$ and for some $j$, then $\psi_i=c$ for all $i$; but then $\ket{\psi}=c\sum_{s=1}^{\infty}\ket{s}$ which does not belong to $\mcal{H}_1$, contrary to the assumption that $\ket{\psi}$ is in $\mcal{H}_1$. Thus if $\braket{i,j}{\psi}=0$ for all $i$ and $j$ for some $\ket{\psi}$ in $\mcal{H}_1$, then $\ket{\psi}=0 $. Thus $\mcal{D}_c^1$ is dense. 

Now we  show that $\mcal{D}_c^1$ is a subspace of $\dom{H_1 T_1}\cap \dom{T_1 H_1}$. Since $\mcal{D}_c^1$ is a proper subspace of $\dom{T_1}$ and $\dom{T_1}$ is invariant under $\opr{H_1}$ (that means $\opr{H_1}:\mcal{D}_c^1$ is a subspace of $\dom{T_1}$), $\opr{T_1 H_1}$ is defined in the entire $\mcal{D}_c^1$. On the other hand, $\dom{T_1}$ is not invariant under $\opr{T}_1$ so that it is not necessary that $\opr{T}_1 \ket{\varphi}$ is in $\dom{H_1}$ for all $\ket{\varphi}$ in $\mcal{D}_c^1$. Now $\dom{H_1 T_1}$ consists of those $\ket{\phi}$ in $\dom{T_1}$ such that $\opr{T_1} \ket{\phi}$ is in $\dom{H_1}$. Specifically the domain consists of those $\ket{\phi}$ in $\dom{T_1}$ such that
\begin{equation}\label{cond}
\sum_{s=1}^{\infty}E_s^2 \abs{\sum_{s'\neq s}^{N} \frac{\varphi_{s'}}{\omega_{s,s'}}}^2<\infty.
\end{equation}
To find the suitable condition to determine the domain of $\opr{H_1 T_1}$, we split the sum in two parts, 
\begin{equation}
\frac{1}{\hbar^2}\sum_{s=1}^{\infty}E_s^2 \abs{\sum_{s'\neq s}^{N}\frac{\varphi_{s'}}{\omega_{s,s'}}}^2 = \sum_{s=1}^{N} E_s^2 \abs{\sum_{s'\neq s}^{N} \frac{\varphi_{s'}}{(E_s-E_{s'})}}^2 + \sum_{s=N+1}^{\infty} E_s^2 \abs{\sum_{s'=1}^{N} \frac{\varphi_{s'}}{(E_s-E_{s'})}}^2.
\end{equation}
For finite $N$ the first term is already finite, so again we need only to concern ourselves with the second term. If the $\ket{\varphi}$'s satisfy the bound
\begin{equation}\label{cond2}
\abs{\sum_{s'=1}^{N} \frac{\varphi_{s'}}{E_s-E_{s'}}}\leq  \frac{C}{E_s^2},
\end{equation}
for some constant $C$ independent of $s$, then $\ket{\varphi}$ belongs to $\dom{H_1 T_1}$.  

It is sufficient then to show that the vectors in $\mcal{D}_c^1$ satisfy (\ref{cond2}) to establish that $\mcal{D}_c^1$ is a proper dense subspace of $\dom{H_1 T_1}$. Now
\begin{eqnarray}
\sum_{s'=1}^{L}\frac{\varphi_{s'}}{E_s-E_{s'}}&=&\sum_{s'=1}^{L}\frac{1}{(E_s-E_{s'})}\left(\sum_{k=1}^{s'-1}a_{s',k}-\sum_{s'+1}^{L}a_{k,s'}\right) \nonumber \\
&=& \frac{1}{E_s^2}\sum_{s'=1}^{L-1}\sum_{k=s'+1}^{L} a_{k,s'}\frac{(E_k-E_{s'})}{(1-E_k/E_s)(1-E_{s'}/E_s)},\nonumber
\end{eqnarray}
where the second line follows from a rearrangement of the first term. The triangle inequality then dictates that
\begin{equation}
\abs{\sum_{s'=1}^{L}\frac{\varphi_{s'}}{E_s-E_{s'}}}
\leq \frac{1}{E_s^2}\sum_{s'=1}^{L-1}\sum_{k=s'+1}^{L} \abs{a_{k,s'}}\frac{(E_k-E_{s'})}{(1-E_k/E_s)(1-E_{s'}/E_s)}.
\end{equation}
Again $(1-E_{s'}/E_s)^{-1}$ is bounded within the range $(L+1)\leq s \leq \infty$ for every $1\leq s' \leq L$. Thus there exists a finite positive constant $B_L$ depending on $L$ alone such that 
\begin{equation}
\sum_{s'=1}^{L-1}\sum_{k=s'+1}^{L} \abs{a_{k,s'}}\frac{(E_k-E_{s'})}{(1-E_k/E_s)(1-E_{s'}/E_s)}< B_L \sum_{s'=1}^{L-1}\sum_{k=s'+1}^{L} \abs{a_{k,s'}}.
\end{equation}
One such $B_L$ is explicitly given by
\begin{equation}
B_L=\max_{1\leq s'\leq (L-1),\, (s'+1)\leq k\leq L \atop (L+1)\leq s\leq \infty} \frac{(E_k-E_{s'})}{(1-E_k/E_s)(1-E_{s'}/E_s)}<\infty.
\end{equation}
Thus we get the required bound of inequality (\ref{cond2})
\begin{equation}
\abs{\sum_{s'=1}^{L}\frac{\varphi_{s'}}{E_s-E_{s'}}}
\leq \frac{1}{E_s^2}\cdot B_L\cdot \sum_{s'=1}^{L-1}\sum_{k=s'+1}^{L} \abs{a_{k,s'}},
\end{equation}
which implies that $\mcal{D}_c^1$ is a subspace of $\dom{H_1 T_1}\cap \dom{T_1 H_1}$. The operators $\opr{H_1}$ and $\opr{T_1}$ are then canonically conjugate in the dense subspace $\mcal{D}_c^1$.

\subsection{$\opr{T}_1$ is Essentially Self-adjoint and It Generates a Class of Essentially Self-adjoint Time Operators}

Now we show that $\opr{T_1}$ is essentially self-adjoint. Let $\opr{T_1}^*$ be the adjoint of $\opr{T_1}$. Then $\opr{T_1}$ in $\dom{T_1}$ is essentially self-adjoint if $\ket{\phi}$ is in $\dom{T_1^*}$ and $(\opr{T_1}^*\pm i\, \opr{I})\ket{\phi}=0$ imply that $\ket{\phi}=0$ (Reed \& Simon 1972). However, it is sufficient to show that $\opr{T_1}$ and its adjoint $\opr{T_1}^*$ are symmetric. This is so because the symmetry of $\opr{T_1}$, with the fact that $\opr{T_1}$ is densely defined, assures the existence of a unique non-trivial adjoint of $\opr{T_1}$, $\opr{T_1}^*$; on the other hand, the symmetry of $\opr{T_1}^*$ dictates that $\opr{T_1}^*$ has real eigenvalues only. If $\opr{T_1}^*$ is symmetric, and if $\ket{\phi}\neq 0$ is in $\dom{T_1^*}$ and $(\opr{T_1}^*\pm i\, \opr{I})\ket{\phi}=0$, then $\ket{\phi}$ is an eigenvector of $\opr{T_1}^*$ with the eigenvalues $\pm i$ which is a contradiction with the assumption that $\opr{T_1}^*$ is symmetric.

First we show that $\opr{T}_1$ is symmetric in its assigned domain $\dom{T_1}$. $\opr{T_1}$ is symmetric if $\braket{\psi}{\opr{T_1}\varphi}=\braket{\opr{T}_1 \psi}{\varphi}$ for all $\ket{\varphi}$, $\ket{\psi}$ in $\dom{T_1}$. Let $\ket{\varphi}=\sum_{s=1}^{N} \varphi_s \ket{s},\,\ket{\psi}=\sum_{s=1}^{L}\psi_s \ket{s}$ in $\dom{T_1}$. Now
\begin{eqnarray}
\braket{\psi}{\opr{T}_1 \varphi}&=&\sum_{s=1}^{L}\psi_{s}^{*} \left(\sum_{s'\neq s}^{N} \frac{i \varphi_{s'}}{\omega_{s,s'}}\right)\nonumber\\
&=&\sum_{s'=1}^{N}\left(\sum_{s\neq s'}^{L} \frac{i \psi_{s}}{\omega_{s',s}}\right)^{*} \varphi_{s'}\nonumber\\
&=&\braket{\opr{T}_1\psi}{\varphi}\nonumber
\end{eqnarray}
where the rearrangement of the summations are possible because they have finite limits. Thus $\opr{T}_1$ is symmetric in its assigned domain.

Now let us determine the adjoint, $\opr{T_1}^*$, of $\opr{T_1}$. Since $\opr{T_1}$ is densely defined and symmetric, it has a unique non-trivial adjoint. The domain $\dom{T_1^*}$ of $\opr{T_1}^*$ consists of those vectors $\ket{\psi}$ in $\mcal{H}_1$ such that there exists a vector $\ket{\psi^*}$ in $\mcal{H}_1$ satisfying the condition $\braket{\psi}{\opr{T_1}\varphi}=\braket{\psi^*}{\varphi}$ for all $\ket{\varphi}$ in $\dom{T_1}$. Let $\ket{\psi}=\sum_{s=1}^{\infty}\psi_s \ket{s}$ be in $\mcal{H}_1$. Then for all $\ket{\varphi}$ in $\dom{T_1}$,
\begin{eqnarray}
\braket{\psi}{\opr{T_1}\varphi}&=&\sum_{s=1}^{\infty} \overline{\psi_s} \sum_{s'\neq s}^{L}\frac{i\varphi_{s'}}{\omega_{s,s'}}\nonumber\\
&=&\sum_{s'=1}^{L}\left(\sum_{s\neq s'}^{\infty} \frac{i\overline{\psi_s}}{\omega_{s,s'}}\right) \varphi_{s'}\nonumber\\
&=&\sum_{s'=1}^{L}\overline{\left(\sum_{s\neq s'}^{\infty} \frac{i\psi_s}{\omega_{s',s}}\right)} \varphi_{s'}\label{3rekrek}.
\end{eqnarray}
Since $L$ is finite, the interchanging of the order of summation in the second line is allowed. With $\psi_{s'}=\sum_{s'\neq s}^{\infty} i\,\omega_{s,s'}^{-1}\,\psi_{s'}$, we can rewrite equation (\ref{3rekrek}) in the form
\begin{eqnarray}
\braket{\psi}{\opr{T_1}\varphi}&=&\sum_{s'=1}^{L}\overline{\psi_{s'}^{*}} \varphi_{s'},\nonumber
\end{eqnarray}
from which we identify the vector 
\begin{equation}\label{3star}
\ket{\psi^{*}}=\sum_{s=1}^{\infty}\left(\sum_{s'\neq s}^{\infty} \frac{i}{\omega_{s,s'}}\,\psi_{s'}\right)\ket{s}
\end{equation}
to satisfy the relation  $\braket{\psi}{\opr{T_1}\varphi}=\braket{\psi^{*}}{\varphi}$ for all $\ket{\varphi}$ in $\dom{T_1}$.
Now $\ket{\psi}$ is in the domain of $\opr{T_1}^*$ if and only if $\ket{\psi^*}$ is in the Hilbert space.  Thus the domain $\dom{T_1^*}$ of $\opr{T_1^*}$ must comprise the following subspace of $\mcal{H}_1$,
\begin{equation}
\dom{T_1^{*}}=\left\{\ket{\psi}=\sum_{s=1}^{\infty}\psi_s \ket{s}\in\mcal{H}_1,\, \sum_{s=1}^{\infty}\abs{\sum_{s'\neq s}^{\infty} \frac{\psi_{s'}}{\omega_{s,s'}}}^2<\infty \right\} .
\end{equation}
Finally the adjoint of $\opr{T_1}$ is uniquely determined by the definition of the adjoint, $\ket{\psi^{*}}=\opr{T_1^{*}}\ket{\psi}$. Equation (\ref{3star}) then yields the adjoint
\begin{equation}
\opr{T}_1^{*}=\sump{s, s'\geq 1}{\infty}\frac{i}{\omega_{s,s}} \ket{s}\bra{s'}.
\end{equation} 
As expected from the symmetry of $\opr{T_1}$, we have the extension relation $\opr{T_1}\subset \opr{T_1}^{*}$. (It may be possible that $\dom{T_1^*}=\dom{T_1}$ already for some systems, in which case $\opr{T_1}$ is immediately self-adjoint.)

To complete the proof that $\opr{T_1}$ is essentially self-adjoint, we now  show that its adjoint $\opr{T_1^{*}}$ is symmetric. It is sufficient to show that for every $\ket{\psi}$ in $\dom{T_1^{*}}$ the number $\braket{\psi}{\opr{T_1^*}\psi}$ is real valued. Let $\ket{\psi}$ be in $\dom{T_1^{*}}$, then
\begin{equation}\label{3sum}
\braket{\psi}{\opr{T_1^*}\psi}=\sum_{s=1}^{\infty}\overline{\psi_s}\sum_{s'\neq s}^{\infty}\frac{i\psi_{s'}}{\omega_{s,s'}}.
\end{equation}
We note that the double sum in equation (\ref{3sum}) is absolutely convergent. This follows from the fact that the first summation is already absolutely convergent, $\ket{\opr{T}_1^*\psi}$ being uniquely determined by $\ket{\psi}$, and
\begin{eqnarray}
\sum_{s=1}^{\infty}\abs{\overline{\psi_s} \sum_{s'\neq s}^{\infty}\frac{\psi_{s'}}{\omega_{s,s'}}}
\leq \sqrt{\sum_{s=1}^{\infty}\abs{\psi_s}^2}\cdot \sqrt{\sum_{s=1}^{\infty}\abs{\sum_{s'\neq s}^{\infty}\frac{\psi_{s'}}{\omega_{s,s'}}}^2}<\infty\nonumber.
\end{eqnarray}
It is well known that if a double sum has been proven to be absolutely convergent for any mode of summation, it will be absolutely convergent for all modes of summation, and the sum of the series is independent of summation (Hardy {\it et al} 1952). Thus we can interchange the order of summation in the right hand side of equation (\ref{3sum}) to give
\begin{eqnarray}
\braket{\psi}{\opr{T_1^*}\psi}&=&\sum_{s'=1}^{\infty}\psi_{s'}\sum_{s\neq s'}^{\infty}\frac{i\overline{\psi_s}}{\omega_{s,s'}}\nonumber\\
&=&\overline{\sum_{s'=1}^{\infty}\overline{\psi_{s'}}\sum_{s\neq s'}^{\infty}\frac{i\psi_s}{\omega_{s',s}}}\nonumber\\
&=&\overline{\braket{\psi}{\opr{T_1^*}\psi}}\nonumber.
\end{eqnarray}
Thus $\opr{T_1^*}$ is symmetric. And $\opr{T_1}$ is consequently essentially self-adjoint.

Now we show that $\opr{T_1}$ generates a class of uncountable essentially self-adjoint time operators conjugate to the same Hamiltonian $\opr{H_1}$. Let $\alpha=\left\{\alpha_{s},\, s=1,2,\dots\right\}$ be a bounded sequence of real numbers, i.e. $\abs{\alpha_s}\leq A<\infty$ for all $s$; these sequences may satisfy some other properties such as $\sum_{s=1}^{\infty}\alpha_s^2 <\infty$, or $\alpha_s=\tau$  for all $s$ for some real $\tau$. Then the operator 
\begin{equation}\label{3baka}
\opr{T_{1,\alpha}}=\opr{T_1}+\sum_{s=1}^{\infty}\alpha_s\ket{s}\!\bra{s}
\end{equation}
is essentially self-adjoint in $\dom{T_1}$. Moreover, $\opr{T_{1,\alpha}}$ is canonically conjugate with $\opr{H_1}$ in $\mcal{D}_c^1$. To prove our assertion, let $\Delta\opr{T_1}=\sum_{s=1}^{\infty}\ket{s}\alpha_{s}\bra{s}$. Obviously $\Delta\opr{T_1}$ is symmetric in $\dom{T_1}$. Thus $\opr{T_{1,\alpha}}$ is essentially self-adjoint in $\dom{T_1}$ if there exists some constants $p_1,\, p_2\geq 0$ such that $\norm{\Delta\opr{T_1}\ket{\varphi}}^2\leq p_1 \braket{\varphi}{\opr{T_1}\varphi} + p_2 \norm{\ket{\varphi}}^2$ for all $\ket{\varphi}$ in $\dom{T_1}$ (Hellwig 1964).  For bounded $\alpha_s$, the operator $\Delta\opr{T_1}$ is bounded; this follows from the fact that $\norm{\Delta\opr{T}_1\ket{\phi}}\leq A \norm{\ket{\phi}}$. Thus there exists some $p_1,\, p_2\geq 0$ such that $\norm{\Delta\opr{T_1}\ket{\varphi}}^2\leq \norm{\Delta\opr{T_1}}^2\norm{\ket{\varphi}}^2\leq p_1 \braket{\varphi}{\opr{T_1}\varphi} + p_2 \norm{\ket{\varphi}}^2$, for all $\ket{\varphi}$ in $\dom{T_1}$; a pair of such $p_1$ and $p_2$ is $p_1=0$ and $p_2=\norm{\Delta\opr{T}_1}$. Therefore $\opr{T_{1,\alpha}}$ is essentially self-adjoint in $\dom{T_1}$. That $\opr{T}_{1,\alpha}$ and $\opr{H_1}$ are canonical in $\mcal{D}_c^1$ follows from the fact that $\opr{T_1}$ are canonical in $\mcal{D}_c^1$, and $\Delta\opr{T_{1}}$ and $\opr{H_1}$ commute in $\mcal{D}_c^1$. 

We note that the above construction of $\Delta\opr{T}_1$ does not necessarilly exhaust all possibilities. However, we emphasize that not all symmetric operators in $\dom{T_1}$ (not even those that are essentially self-adjoint or self-adjoint in $\dom{T_1}$) which commute with the Hamiltonian in $\mcal{D}_c^1$ can be added to $\opr{T}_1$ to give an essentially self-adjoint time operator in $\dom{T_1}$. If $\opr{T}_1$ is bounded and if $\Delta\opr{T}_1$ is unbounded in $\dom{T_1}$, then no constants $p_1,p_2\geq 0$ can satisfy $\norm{\Delta\opr{T_1}\ket{\varphi}}^2\leq p_1 \braket{\varphi}{\opr{T_1}\varphi} + p_2 \norm{\ket{\varphi}}^2$ for all $\ket{\varphi}$ in $\dom{T_1}$ because for every $p_1$ and $p_2$ there will always be a $\ket{\psi}$ in $\dom{T_1}$ such that $\norm{\Delta\opr{T_1}\ket{\psi}}^2\geq  p_1 \braket{\psi}{\opr{T_1}\psi} + p_2 \norm{\ket{\psi}}^2$, $\Delta \opr{T}_1$ being unbounded and the right hand side of the inequality being always bounded for bounded $\opr{T}_1$.

\section{Characteristic Time Operators for\\M-Degenerate Hamiltonians}\label{deg}

Now let us consider the case when the Hamiltonian has a constant degeneracy $M$, i.e. to every energy eigenvalue $E_s$ corresponds to $M$ linearly independent and orthonormal eigenvectors, $\ket{s,r}$, where $\braket{s,r}{s',r'}=\delta_{ss'}\delta_{rr'}$, in which $r=1,\,,\dots,\, M$  and $s=1,\,2,\,\dots$. We assume that the Hilbert space is spanned by these eigenvectors,
\begin{equation}
\mcal{H}_M=\left\{\ket{\varphi}=\sum_{s=1}^{\infty}\sum_{r=1}^M \varphi_{s,r}\ket{s,r},\, \sum_{n=1}^{\infty}\sum_{r=1}^M \abs{\varphi_{s,r}}^2< \infty\right\}
\end{equation}
and $\mcal{H}_M$ is equipped with the standard norm $\norm{\ket{\varphi}}=\sqrt{\sum_{n=1}^{\infty}\sum_{r=1}^M \abs{\varphi_{s,r}}^2}$ derived from the standard inner product $\braket{\psi}{\phi}=\sum_{n=1}^{\infty}\sum_{r=1}^M \psi_{n,r}^{*} \phi_{n,r}$. Under this representation, the Hamiltonian $\opr{H}_M$ with domain $\dom{H_M}$ is explicitly given by
\begin{equation}
\opr{H_M}=\sum_{s=1}^{\infty}\sum_{r=1}^M E_s \ket{s,r} \! \bra{r,s},
\end{equation}
\begin{equation}
\dom{H_M}=\left\{\ket{\varphi}=\sum_{s=1}^{\infty}\sum_{r=1}^M \varphi_{s,r}\ket{s,r}\in\hilbert,\,\sum_{s=1}^{\infty}\sum_{r=1}^M E_s^2 \abs{\varphi_{s,r}}^2<\infty\right\}.
\end{equation}

Now given the eigenvectors $\ket{s,r}$ and eigenvalues $E_s$ of the Hamiltonian, we construct the following formal symmetric operator,
\begin{equation}
\opr{T}_M=\sump{s, s'\geq 1}{\infty}\sump{r, r'\geq 1}{M}\frac{i}{\omega_{s,s'}}\ket{s,r}\bra{r',s'},
\end{equation}
where $\omega_{s,s'}=(E_s-E_{s'})/\hbar$, and again the primes indicate that $s=s'$ and $r=r'$ are excluded from the summation. Similarly, $\oprM{T}$ has a dimension of time and it is solely constructed out of the spectral decomposition of the Hamiltonian alone. Similarly we will show below that if the eigenvalues of the Hamiltonian satisfies the condition
\begin{equation}\label{3conconcon}
\sum_{s=1}^{\infty}\frac{1}{E_s^2}<\infty,
\end{equation}
then the formal operator $\opr{T_M}$ can be assigned a dense subspace in $\mcal{H}_M$, and it is canonically conjugate with the Hamiltonian in some dense subspace of $\mcal{H}_M$, and it is essentially self-adjoint in its assigned domain. Likewise we call $\opr{T}_M$ as the characteristic time operator for the M-degenerate Hamiltonian.

\subsection{$\opr{T}_M$ is Densely Definable}

Now we show that if condition (\ref{3conconcon}) is satisfied,
then $\oprM{T}$ can be assigned the following dense subspace, 
\begin{equation}
\mcal{D}_M=\left\{\ket{\varphi}=\sum_{s=1}^N \sum_{r=1}^M \varphi_{s,r} \ket{s,r},\, \abs{\varphi_{s,r}}<\infty,\, N<\infty\right\},
\end{equation}
and in this subspace it is essentially self-adjoint. The subspace $\mcal{D}_M$ is dense because for every $\ket{\psi}$ in $\mcal{H}_M$ the sequence of elements in $\mcal{D}_M$ given by 
\begin{displaymath}
\left\{\ket{\phi_k}=\sum_{s=1}^{k}\sum_{r=1}^{M}\braket{s,r}{\psi}\ket{s,r},\,k=1,\, 2,\, \dots\right\}
\end{displaymath}
converges to $\ket{\psi}$.

We first show that the formal time operator $\opr{T}_M$ is defined in the entire $\mcal{D}_M$; that is, $\opr{T}_M\ket{\varphi}$ is in $\mcal{H}_M$ for all $\ket{\varphi}$ in $\mcal{D}_M$. Let $\ket{\varphi}$ be in $\mcal{D}_M$, then 
\begin{equation}
\opr{T}_M\ket{\varphi}=\sum_{s=1}^{\infty} \sum_{r=1}^M \left(\sum_{s'\neq s}^{N} \sum_{r'\neq r}^M \frac{i\varphi_{s',r'}}{\omega_{s,s'}}\right) \, \ket{s,r}.
\end{equation}
$\opr{T}_M\ket{\varphi}$ lies in the Hilbert space $\hilbert$ if and only if $\norm{\opr{T}_M\ket{\varphi}}<\infty$, or equivalently,
\begin{equation}
\norm{\opr{T}_M\ket{\varphi}}^2=\sum_{s=1}^{\infty}\sum_{r=1}^M \abs{\sum_{s'\neq s}^{N} \sum_{r'\neq r}^M \frac{\varphi_{s',r'}}{\omega_{s,s'}}}^2<\infty. 
\end{equation}
We divide the sum in two parts,
\begin{eqnarray}\label{ggg}
\frac{1}{\hbar^2} \sum_{s=1}^{\infty}\sum_{r=1}^M \abs{\sum_{s'\neq s}^{N}\sum_{r'\neq r}^M \frac{\varphi_{s'}}{\omega_{s,s'}}}^2&=& \sum_{s=1}^{N}\sum_{r=1}^M \abs{\sum_{s'\neq s}^{N} \sum_{r'\neq r}^M \frac{\varphi_{s',r'}}{(E_s-E_{s'})}}^2\nonumber\\
& &\hspace{1.5mm} + \sum_{s=N+1}^{\infty}\sum_{r=1}^M \abs{\sum_{s'=1}^{N}\sum_{r'\neq r}^M \frac{\varphi_{s',r'}}{(E_s-E_{s'})}}^2.
\end{eqnarray}
For finite $N$ the first term is already finite, so we need only to show that the second term is finite. 

Again appealing to the boundedness of $(1-E_{s'}/E_s)^{-1}$ within the range $(N+1)\leq s \leq \infty$ for all $s'<N$, we have the following bound,
\begin{eqnarray}
\sum_{r=1}^M \abs{\sum_{s'=1}^{N}\sum_{r'\neq r}^M \frac{\varphi_{s',r'}}{(E_s-E_{s'})}}^2&\leq &  \sum_{r=1}^M \left(\sum_{s'=1}^{N}\sum_{r'\neq r}^M \frac{\abs{\varphi_{s',r'}}}{(E_s-E_{s'})}\right)^2 \nonumber\\
&\leq &\frac{1}{E_s^2}\cdot A_N^2 \cdot \sum_{r=1}^M \left(\sum_{s'=1}^{N}\sum_{r'\neq r}^M \abs{\varphi_{s',r'}}\right)^2. \label{kawawa}
\end{eqnarray}
The second term then takes the bound
\begin{equation}
\sum_{s=N+1}^{\infty}\sum_{r=1}^M \abs{\sum_{s'=1}^{N}\sum_{r'\neq r}^M \frac{\varphi_{s',r'}}{(E_s-E_{s'})}}^2 \leq  A_N^2 \cdot \sum_{r=1}^M \left(\sum_{s'=1}^{N}\sum_{r'\neq r}^M \abs{\varphi_{s',r'}}\right)^2 \cdot \sum_{s=N+1}^{\infty}\frac{1}{E_s^2},
\end{equation}
which is finite as the sum $\sum_{s}E_s^{-2}<\infty$ is finite by assumption. Since $\ket{\varphi}$ is an arbitrary element of $\mcal{D}_M$, the bound (\ref{kawawa}) holds for every element of $\mcal{D}_M$. Thus $\opr{T}_M\ket{\varphi}$ is in $\mcal{H}_M$ for all $\ket{\varphi}$ in $\mcal{D}_M$.  Since $\opr{T}_M$ is densely defined in $\mcal{D}_M$, we define $\opr{T}_M$ as the densely defined operator $\opr{T}_M:\mcal{D}_M\subseteq \mcal{H}_M\mapsto \mcal{H}_M$ with domain $\mcal{D}(\opr{T}_M)=\mcal{D}_M$. 

\subsection{$\opr{T}_M$ and $\opr{H}_M$ are Canonically Conjugate}

Now we show that the pair of operators $\opr{T}_M$ and $\opr{H}_M$  form a canonical pair in a dense subspace $\mcal{D}_c^M$ of $\mcal{D}(\opr{H}_M \opr{T}_M)\cap \dom{T_2 H_M}$, i.e. $(\oprM{T}\oprM{H}-\oprM{H}\oprM{T})\ket{\varphi}=i\hbar \ket{\varphi}$ for all $\ket{\varphi}$ in $\mcal{D}_c^M$. Again we assume for the moment that $\domMM{H}{T}\cap \domMM{T}{H}$ is not empty. If $\ket{\varphi}$ is in $\domMM{H}{T}\cap \domMM{T}{H}$, then 
\begin{equation}
\left(\oprM{T} \oprM{H}-\oprM{H}\oprM{T}\right)\ket{\varphi}=-i\hbar\sum_{s=1}^{\infty} \sum_{r=1}^{M}\left(\sum_{s'\neq s}^{L}\sum_{r'\neq r}^{M}\varphi_{s',r'}\right)\ket{s,r}.\label{cancancan}
\end{equation}
If 
\begin{equation}\label{conconcon}
\sum_{s=1}^{L}\sum_{r=1}^{M}\varphi_{s,r}=0,
\end{equation}
then $\sum_{s'\neq s}^{L}\sum_{r'\neq r}^{M}\varphi_{s'}=-\varphi_{s,r}$. In which case equation (\ref{cancancan}) reduces to 
\begin{equation}
\left(\oprM{T} \oprM{H}-\oprM{H}\oprM{T}\right)\ket{\varphi}=i\hbar\ket{\varphi}.
\end{equation}
That is $\oprM{T}$ and $\oprM{H}$ satisfy the canonical commutation relation in the subspace of vectors satisfying equation (\ref{conconcon}). 

We have then to identify a dense subspace of $\domM{T}$ satisfying (\ref{conconcon}) and at the same time belonging to $\domMM{H}{T}\cap \domMM{T}{H}$. Now we show that the vectors of the following proper subspace of $\domM{T}$ satisfy these conditions,
\begin{equation}
\mcal{D}_c^{M}=\left\{\ket{\xi}=\sum_{j=1}^{L-1} \sum_{i=j+1}^L  \sum_{k=1}^M a_{i,j,k} \left(\ket{i,k}-\ket{j,k}\right),\, a_{i,j,k}\in \mcal{C},\, \mbox{for finite even}\, L>1\right\}.
\end{equation}
In expanded form the vectors become
\begin{equation}
\ket{\varphi}=\sum_{k=1}^{M}\sum_{s=1}^{L}\varphi_{s,k}\ket{s,k}=\sum_{k=1}^{M}\sum_{s=1}^{L}
\left(\sum_{l=1}^{s-1}a_{s,l,k}-\sum_{l=s+1}^{L}a_{l,s,k}\right)\ket{s,k},
\end{equation}
provided we set $a_{i,j,k}=0$ when either $i$ or $j$ is outside of its respective range, $1\leq j\leq (L-1)$, $(j+1)\leq i\leq L$; or when both are outside of their ranges; or when $i\leq j$. Similar calculation shows that the sum of the coefficients for each $k$ vanishes. Thus the vectors in $\mcal{D}_c^{M}$ satisfy condition (\ref{conconcon}).  Thus to establish our claim, it is sufficient to show that $\mcal{D}_c^M$ is dense and that $\mcal{D}_c^M$ is a subspace of $\domMM{H}{T}\cap \domMM{T}{H}$. 

Now $\mcal{D}_c^{M}$ is dense. Let us assume otherwise. Then there exists a vector $\ket{\psi}$ in $\mcal{H}_M$ which is not the zero vector such that $\braket{\xi}{\psi}=0$ for all $\ket{\xi}$ in $\mcal{D}_c^M$. Since $\ket{i,j,k}=(\ket{i,k}-\ket{j,k})$ is in $\mcal{D}_c^M$ for all pair of $i$ and $j$, and for all $k=1,\,\dots M$. Then $\braket{i,j,k}{\psi}=0$ implies $\psi_{i,k}-\psi_{j,k}=0$ or $\psi_{i,k}=\psi_{j,k}$. Let us say that for all $k$ there exists some $j$ (not necessarilly the same $j$ for each $k$) such that $\psi_{j,k}=0$, which implies that $\ket{\psi}=0$, a contradiction with the assumption that $\ket{\psi}\neq 0$.  If on the other hand $\psi_{j,k}$ is equal to some finite constant $c_k$ for some $j$ for a particular $k$, then $\psi_{i,k}=c_k$ for all $i$; but for this case, the vector will be given by $\ket{\psi}=\sum_{s=1}^{\infty}\sum_{r=1}^M c_r \ket{s,r}$ which does not lie in the Hilbert space, a contradiction with the assumption that $\ket{\psi}$ is in $\mcal{H}_M$. Then if $\braket{\xi}{\psi}=0$ for all $\ket{\xi}$ in $\mcal{D}_c^M$ we must have $\ket{\psi}=0 $. Thus $\mcal{D}_c^M$ is dense. 

To complete the proof, now we show that $\mcal{D}_c^{M}$ is a subspace of $\domMM{H}{T}\cap \domMM{T}{H}$. Since $\mcal{D}_c^M$ is a subset of $\domM{T}$ and $\domM{T}$ is invariant under $\oprM{H}$, $\oprM{T}\oprM{H}$ is defined in the entire $\mcal{D}_c^{M}$. Now $\domMM{H}{T}$ consists of those $\ket{\phi}$ in $\domM{T}$ such that $\oprM{T} \ket{\phi}$ is in $\domM{H}$. Specifically the domain consists of those $\ket{\phi}$ such that
\begin{equation}\label{rere}
\norm{\opr{T}_M\ket{\phi}}^2=\sum_{s=1}^{\infty}E_s^2 \sum_{r=1}^M  \abs{\sum_{s'\neq s}^{L} \sum_{r'\neq r}^M \frac{\phi_{s',r'}}{\omega_{s,s'}}}^2<\infty.
\end{equation}
Let us  divide equation (\ref{rere}) in two parts,
\begin{eqnarray}
\frac{1}{\hbar^2}\sum_{s=1}^{\infty}E_s^2 \sum_{r=1}^M  \abs{\sum_{s'\neq s}^{L} \sum_{r'\neq r}^M \frac{\phi_{s',r'}}{\omega_{s,s'}}}^2&=&\sum_{s=1}^{L}E_s^2 \sum_{r=1}^M  \abs{\sum_{s'\neq s}^{L} \sum_{r'\neq r}^M \frac{\phi_{s',r'}}{(E_s-E_{s'})}}^2\nonumber\\
& &\hspace{1.5mm} + 
\sum_{s=L+1}^{\infty}E_s^2 \sum_{r=1}^M  \abs{\sum_{s'=1}^{L} \sum_{r'\neq r}^M \frac{\phi_{s',r'}}{(E_s-E_{s'})}}^2.
\end{eqnarray}
The first term is already finite for finite $L$ so that equation (\ref{rere}) is satisfied as long as the second term is finite. Now all of $\mcal{D}_c^{M}$ is in $\domMM{H}{T}\cap \domMM{T}{H}$ if
\begin{equation}\label{leklek}
\abs{\sum_{s'=1}^{L} \sum_{r'\neq r}^M \frac{\phi_{s',r'}}{(E_s-E_{s'})}}\leq \frac{D}{E_s^2}
\end{equation}
for some positive finite $D$ independent of $s$. 

Now we show that the vectors $\ket{\phi}$ in $\mcal{D}_c^{M}$ satify this condition. For a given $\rho$, we have
\begin{equation}\label{ooo}
\sum_{s'=1}^L \frac{\phi_{s',\rho}}{E_s-E_{s'}}=\sum_{i=1}^{L-1}\sum_{j=i+1}^{L}a_{j,i,\rho} \frac{(E_j-E_i)}{(E_s-E_i)(E_s-E_j)}
\end{equation}
for $s>L$, where we have used the same method of rearrangement to arrive at equation (\ref{ooo}). Again we appeal to the boundedness of $(1-E_{s'}/E_s)^{-1}$ within the indicated range to get the bound
\begin{eqnarray}
\abs{\sum_{s'=1}^L \frac{\phi_{s',\rho}}{E_s-E_{s'}}}&\leq& \frac{1}{E_s^2}\sum_{i=1}^{L-1}\sum_{j=i+1}^{N}\abs{a_{j,i,\rho}} \frac{(E_j-E_i)}{(1-E_i/E_s)(1-E_j/E_s)}\nonumber\\
&\leq  &\frac{1}{E_s^2}\cdot B_L \cdot \sum_{i=1}^{L-1}\sum_{j=i+1}^{N}\abs{a_{j,i,\rho}}.\label{www}
\end{eqnarray}
Comparing this with equation (\ref{leklek}), we find that every $\ket{\psi}$ in $\mcal{D}_c^M$ belongs to $\domMM{H}{T}$, which implies that $\mcal{D}_c^{M}$ is a subspace of $\domMM{H}{T}$. Thus the operators $\opr{H}_M$ and $\opr{T}_M$ are canonically conjugate in the dense subspace $\mcal{D}_c^M$.

\subsection{$\opr{T}_M$ is Essentially Self-adjoint and It Generates a Class of Essentially Self-adjoint Time Operators}

Finally we prove that $\oprM{T}$ is essentially self-adjoint in its assinged subspace. First we show that $\opr{T}_M$ is symmetric in $\mcal{D}$. Let $\ket{\varphi}=\sum_{s=1}^{N}\sum_{r=1}^{M}\varphi_{s,r}\ket{s,r}$ and $\ket{\phi}=\sum_{s=1}^{L}\sum_{r=1}^{M}\phi_{s,r}\ket{s,r}$, then
\begin{eqnarray}
\braket{\psi}{\opr{T}_M \varphi}&=& \sum_{s=1}^{L} \sum_{r=1}^{M}\psi_{s,r}^{*}\left(\sum_{s'\neq s}^{N} \sum_{r\neq r'}^{M} \frac{i \varphi_{s',r'}}{\omega_{s,s'}}\right)\nonumber\\
&=& \sum_{s'=1}^{N} \sum_{r'=1}^{M} \left(\sum_{s\neq s'}^{N} \sum_{r\neq r'}^{M} \frac{i \psi_{s',r'}}{\omega_{s',s}}\right)^{*} \varphi_{s,r}\nonumber\\
&=&\braket{\opr{T}_M \psi}{\varphi},\nonumber
\end{eqnarray}
where the rearrangement is possible because the limits of summations are finite. Thus $\opr{T}_M$ is symmetric.

Since $\opr{T}_M$ is densely defined and symmetric, it is assured that it has a unique adjoint $\opr{T}_M^*$.  Let $\ket{\phi}$ be in $\mcal{D}(\opr{T}_M^*)$. Then for all $\ket{\varphi}$ in $\domM{T}$, 
\begin{eqnarray}
\braket{\phi}{\opr{T}_M\varphi}&=&\sum_{s=1}^{\infty} \sum_{r=1}^{M} \overline{\phi_{s,r}} \sum_{s'\neq s}^{L}\sum_{r'\neq r}^{M}\frac{i\varphi_{s',r'}}{\omega_{s,s'}}\nonumber \\
&=&\sum_{s'=1}^N \sum_{r'=1}^M \left(\sum_{s\neq s'}^{\infty}\sum_{r\neq r'}^{M} \frac{i\overline{\phi_{s,r}}}{\omega_{s,s'}}\right) \, \varphi_{s'r'}\nonumber\\
&=&\braket{\phi^*}{\varphi}\nonumber
\end{eqnarray}
where
\begin{equation}
\ket{\phi^*}=\sum_{s=1}^{\infty}\sum_{r=1}^{M}\left(\sum_{s'\neq s}^{\infty}\sum_{r'\neq r}^{M} \frac{i\phi_{s',r'}}{\omega_{s,s'}}\right)\, \ket{s,r}\label{3rek}
\end{equation}
It must be that $\ket{\psi^*}$ is in the Hilbert space. Thus the domain of $\opr{T}_M^*$ consists of the following subspace of $\mcal{H}_M$,
\begin{equation}
\mcal{D}\left(\opr{T}_M^*\right)=\left\{\ket{\phi}=\sum_{s=1}^{\infty}\sum_{r=1}^M \phi_{s,r}\ket{s,r}\in\mcal{H},\, \sum_{s=1}^{\infty}\sum_{r=1}^{M}\abs{\sum_{s'\neq s}^{\infty}\sum_{r'\neq r}^{M} \frac{{\phi_{s',r'}}}{\omega_{s,s'}}}^2<\infty\right\}
\end{equation}
And since the adjoint is defined by $\ket{\phi^*}=T_M^*\ket{\phi}$, it can be directly extracted from equation (\ref{3rek}) to give
\begin{equation}\label{3mad}
\opr{T}_M^*=\sump{s, s' \geq 1}{\infty}\sump{r, r' \geq 1}{M}\frac{i}{\omega_{s,s'}}\ket{s,r}\bra{r',s'}.
\end{equation}
Instead of showing that $\opr{T}_M^*$ is symmetric to prove the essential self-adjointness of $\opr{T}_M$, we demonstrate that if $\opr{T}_M$ is not essentially self-adjoint then we can contradict our earlier conclusion on the non-degenerate case. This is to show that the non-degenerate and degenerate cases are intimately related. (It may be possible that $\mcal{D}\left(\opr{T}_M^*\right)=\mcal{D}\left(\opr{T}_M\right)$ already for some systems, in which case $\opr{T}_M$ is immediately self-adjoint.)

Let us assume the contrary that $\opr{T}_M$ is not essentially self-adjoint. Then there exists a vector $\ket{\eta}=\sum_{s=1}^{\infty}\sum_{r=1}^M \eta_{s,r}\ket{s,r}$ in $\mcal{D}\left(\opr{T}_M^*\right)$ which is not the zero vector such that 
\begin{equation}\label{3man}
\left(\opr{T}_M^*\pm i\,\opr{I}\right)\ket{\eta}=0.
\end{equation}
(We can always rescale in units such that we don't need to be bothered with units in (\ref{3man}).) Inserting the explicit expansion of $\ket{\eta}$ and using the representation of $\opr{T}_M^*$ given by (\ref{3mad}) in equation (\ref{3man}), yield
\begin{equation}\label{3ekek}
\sum_{s\neq \sigma}^{\infty}\frac{i}{\omega_{s,\sigma}} \sum_{r\neq \rho}^{M} \eta_{s,r}=\pm i\, \eta_{\sigma,\rho}.
\end{equation}
for every $\sigma$ and $\rho$. Since $\ket{\eta}$ is asserted to exist, the summation in (\ref{3ekek}) must converge absolutely for all $\sigma$ and for every $\rho$, including the limit $\sigma\to \infty$. Now for two different values of $\rho$, say $k$ and $l$, we get the two coupled equations,
\begin{eqnarray}
\sum_{s\neq \sigma}^{\infty}\frac{i}{\omega_{\sigma,s}} \eta_{s,k}+\sum_{s\neq \sigma}^{\infty}\frac{i}{\omega_{s,\sigma}} \sum_{r\neq k,l}^{M}\eta_{s,r}&=&\pm i\,  \eta_{\sigma,l}\label{11}\\
\sum_{s\neq \sigma}^{\infty}\frac{i}{\omega_{\sigma,s}} \eta_{s,l}+\sum_{s\neq \sigma}^{\infty}\frac{i}{\omega_{s,\sigma}}\sum_{r\neq l,k}^{M} \eta_{s,r}&=&\pm i\, \eta_{\sigma,k}\label{22}
\end{eqnarray}
Subtracting equation (\ref{22}) from equation (\ref{11}) gives
\begin{equation}\label{kawkaw}
\sum_{s\neq \sigma}^{\infty}\frac{i}{\omega_{s,\sigma}}\left(\eta_{s,k}-\eta_{s,l}\right)=\mp i\, \left(\eta_{\sigma,k}-\eta_{\sigma,l}\right).
\end{equation}
Since $\ket{\psi}$ belongs to the Hilbert space, we must have ${\sum_{\sigma=1}^{\infty}\abs{\left(\eta_{\sigma,k}-\eta_{\sigma,l}\right)}^2}<\infty$. 

Now where is the contradiction? We note that $\mcal{H}_M$ is a direct sum of $M$ Hilbert spaces spanned separately by the $M$ degenerate eigenvectors of the Hamiltonian $\opr{H}_M$, i.e. $\mcal{H}_M=\mcal{H}_1\oplus \mcal{H}_2\oplus \dots \oplus \mcal{H}_M$, where $\mcal{H}_{\rho}$ for all  $1\leq \rho\leq M$ is spanned by $\ket{s,\rho}$, $s=1,\,2,\,\dots$.  Let $\opr{P}_{\rho}$ be the projection operator unto the subspace $\mcal{H}_{\rho}$. Then each of the subspaces $\mcal{D}_{\rho}=\opr{P}_{\rho}\dom{H_M}$,  $\rho=1,\,2\,\dots,M$, is invariant under the Hamiltonian; and these subspaces consequently reduce the Hamiltonain.  Then the restriction $\opr{H}_{M_{\rho}}=\opr{P}_{\rho}\opr{H}_M \opr{P}_{\rho}$ of the Hamiltonian $\opr{H}_M$ on $\mcal{H}_{\rho}$ is a self-adjoint, nondegerate Hamiltonian on $\mcal{H}_{\rho}$ with the same eigenvalues $E_s$ as the eigenvalues of the original Hamiltonain $\opr{H}_M$. Since $\sum_{s=1}^{\infty}E_s^{-2}<\infty$, for some fixed $\rho$ we can construct the following characteristic time operator for the non-degenerate Hamiltonian $\opr{H}_{M_{\rho}}$ on $\mcal{H}_{\rho}$,
\begin{equation}
\opr{T}_1^{\rho}=\sump{s, s'\geq 1}{\infty}\frac{i}{\omega_{s,s'}}\ket{s,\rho}\bra{s',\rho},\nonumber
\end{equation}
\begin{equation}
\dom{T_1^{\rho}}=\left\{\ket{\psi}=\sum_{s=1}^{N}\psi_s \ket{s,\rho},\, N<\infty\right\}.\nonumber
\end{equation}
$\opr{T}_1^{\rho}$ has all the properties as those considered in the non-degenerate case; and it is, most importantly, essentially self-adjoint. Thus there exists no $\ket{\phi}=\sum_{s=1}^{\infty}\phi_s \ket{s,\rho}$ in $\dom{{T_1^{\rho}}^*}$ different from the zero vector such that $(\opr{T_M^{\rho}}^* \pm i\, \opr{I})\ket{\phi}=0$. Had such a vector existed, then its coefficients $\phi_s$'s, not all vanishing, would have satisfied the equation
\begin{equation}\label{3cond3}
\sum_{s\neq \sigma}^{\infty} \frac{i}{\omega_{s,\sigma}} \phi_s=\pm i\, \phi_{\sigma}.
\end{equation}
However, if $\opr{T}_M$ is not essentially self-adjoint then it is possible that $\eta_{s,k}-\eta_{s,l}\neq 0$ for some $s$, which implies that (\ref{3cond3}) has a non-trivial solution since we can choose $\phi_s=\eta_{s,k}-\eta_{s,l}$, owing from the fact that ${\sum_{\sigma=1}^{\infty}\abs{\left(\eta_{\sigma,k}-\eta_{\sigma,l}\right)}^2}<\infty$. But this is in contradiction with the essential self-adjointness of $\opr{T_1}^{\rho}$.

Therefore in order to maintain the equality in equation (\ref{kawkaw}) and avoid any contradiction, we must have $\eta_{s,k}=\eta_{s,l}$ for all $k$ and $l$ for every $s$, i.e. $\eta_{s,1}=\eta_{s,2}=\dots=\eta_{s,M}$. Then for all $\rho$, equation (\ref{3ekek}) reduces to
\begin{equation}\label{3wawa}
\sum_{s\neq \sigma}^{\infty}\frac{i(M-1)}{\omega_{s,\sigma}} \eta_{s,\rho}=\pm i\, \eta_{\sigma,\rho}.
\end{equation}
But equation (\ref{3wawa}) is equivalent to equation (\ref{3cond3}) only that the Hamiltonian is scaled to $\opr{H}_1^{\rho}/(M-1)$. Again because $\ket{\eta}$ belongs to the Hilbert space which implies $\sum_{s=1}^{\infty}\abs{\eta_{s,\rho}}^2<\infty$, the assertion that $\opr{T}_M$ is not essentially self-adjoint implies that the charateristic time operator corresponding to (\ref{3wawa}) is not essentially self-adjoint, which is a contradiction with our earlier result for non-degenerate Hamiltonians. Thus  we must necessarilly have $\eta_{s,r}=0$ for all $s$ and $r$, implying that $\ket{\eta}$ is the zero vector, contrary to the assumption that it is otherwise. Therefore  $\opr{T}_M$ must be essentially self-adjoint in its assigned domain.

As in the non-degenerate case, the characteristic time operator $\oprM{T}$ generates a class of uncountable essentially self-adjoint operator canonically conjugate with the same Hamilotnian $\oprM{H}$.  Now let $\alpha=\left\{\alpha_{s,r},\, s=1,2,\dots,\, 1\leq r\leq M\right\}$ be a bounded sequence of real numbers in $s$ for every $r$, i.e. $\abs{\alpha_{s,r}}<A<\infty$ for all $s$ and $r$. Then the operator 
\begin{equation}
\opr{T}_{M,\alpha}=\oprM{T}+\sum_{s=1}^{\infty}\sum_{r=1}^M \alpha_{s,r}\ket{s,r}\! \bra{s,r}
\end{equation}
is essentially self-adjoint in $\domM{T}$. Moreover, $\opr{T}_{M,\alpha}$ is canonically conjugate with $\oprM{H}$ in $\mcal{D}_c^M$. We can follow the same line of proof as the one we used in the non-degenerate case to prove our assertation.

We note that the argument found in (Pegg 1998) and (Jordan 1927) on the non-existence of self-adjoint time operators for discrete Hamiltonians assumes that the eigenvectors of the Hamiltonian belong to the domain $\mcal{D}_c^M$ for all $M\geq 1$. The eigenvectors, however, lie outside of $\mcal{D}_c^M$, so that no contradiction arises.

\section{Conclusion}
The essential self-adjointness of $\opr{T}_M$ (for all $1\leq M<\infty$) means that there exists a unique self-adjoint operator $\overline{\opr{T}}_M:\mcal{D}(\overline{\opr{T}}_M)\subseteq \mcal{H}\mapsto \mcal{H}$  whose reduction in $\domM{T}$ is $\opr{T}_M$ itself. Equivalently $\overline{\opr{T}}_M$ is the unique self-adjoint extention of $\opr{T}_M$. Similarly, for every bounded sequence $\alpha$, there exists a unique self-adjoint operator $\overline{\opr{T}}_{M,\alpha}:\mcal{D}(\overline{\opr{T}}_{M,\alpha})\subseteq \hilbert\mapsto \hilbert$ whose reduction in $\domM{T}$ is $\opr{T_{M,\alpha}}$. Because $\overline{\opr{T}}_M$ and $\overline{\opr{T}}_{M,\alpha}$ are extensions of  $\opr{T}_M$ and $\opr{T}_{M,\alpha}$, respectively, they remain canonically conjugate with the Hamiltonian $\opr{H}_M$ in the same dense subspace $\mcal{D}_c^{(M)}$. The self-adjoint operators $\overline{\opr{T}}_{M}$ and $\overline{\opr{T}}_{M,\alpha}$ are just the adjoints $\opr{T_M}^*$ and $\opr{T_{M,\alpha}}^*$, respectively. Thus, in this paper, we have explicitly proven the following 
\begin{theorem}\label{ttt}
Given a self-adjoint Hamiltonian $\opr{H}$ possessing the following properties:
\begin{enumerate}
\item It has a pure point spectrum bounded from below which can be ordered according to size,  i.e. $-\infty< E_1< E_2<E_3<\cdots$,
\item It has a constant finite degeneracy $1\leq M<\infty$,
\item The sum of the reciprocal of the square of its eigenvalues is finite, i.e $\sum_{s=1}^{\infty}E_{s}^{-2}<\infty$,
\item Its eigenvectors span the entire Hilbert space.
\end{enumerate}
Then there exists a self-adjoint time operator $\opr{T}$ characteristic of the system which is canonically conjugate with the Hamiltonian in a dense subspace of $\dom{TH}\cap\dom{HT}$, i.e. 
\begin{equation}
\left(\opr{TH}-\opr{HT}\right)\subset i\hbar \opr{I},\nonumber
\end{equation}
where $\opr{I}$ is the identity of the Hilbert space $\mcal{H}$. Moreover, $\opr{T}$ generates a class of uncountably many other self-adjoint time operators canonically conjugate with the same Hamiltonian in the same dense proper subspace of the Hilbert space.
\end{theorem}

One might notice that the above theorem implicitly assumes that the Hamiltonian has no zero eigenvalue. But the above theorem can be extended without difficulty to Hamiltonians with zero eigenvalues. One only needs to modify condition (3) to require {\it that the sum of the reciprocal of the non-vanishing eigenvalues is finite}.The construction of the characteristic time operator and the class it generates are the same as in the cases considered here. We mention that the bounded and self-adjoint operator canonically conjugate with the number operator constructed by Garisson and Wong (1970), and Galindo (1984) is an example (See also Busch 1995{\it a},{\it b}).

To conclude, we give a class of characteristic time operators. Consider the class of characteristic time operators distinguished by the following requirement further imposed upon the eigenvalues of the Hamiltonian,
\begin{equation}\label{compact}
{\sump{s, s' \geq 1}{\infty}}\frac{1}{\left(E_s-E_{s'}\right)^2}<\infty.
\end{equation}
We note that when (\ref{compact}) is satisfied, the condition $\sum_{s=1}^{\infty}E_s^{-2}<\infty$ is automatically satisfied. Under this condition, the characteristic essentially self-adjoint time operator for both degenerate and non-degenerate Hamiltonians admits a bounded and compact self-adjoint extension. The boundedness of $\opr{T}_M$ for every finite $M\geq 1$ follows from the following inequality, 
\begin{equation}
\norm{\opr{T}_M\ket{\varphi}}\leq M \left({\sump{s, s' \geq 1}{\infty}}\frac{1}{\omega_{s,s'}^2}\right)^{\frac{1}{2}}\norm{\ket{\varphi}}
\end{equation}
for every $\ket{\varphi}$ in $\mcal{D}(\opr{T}_M)$. The boundedness of $\opr{T}_M$ means that it can be extended in the entire Hilbert space. And since $\opr{T}_M$ is symmetric, it is self-adjoint in the entire $\mcal{H}$.

The compactness of $\opr{T}_M$ for every finite $M\geq 1$ can be shown in the configuration space representation in which $\opr{T}_M$ assumes the form of a Fredholm integral operator with square integrable kernel. That is for every $\varphi(q)=\braket{q}{\varphi}$ in the domain of $\opr{T}_M$ in configuration space representation, we have
\begin{equation}
\left(\opr{T}_M\varphi\right)\!(q)=\int_{\Omega}\kernelM{T}{M}\varphi(q')\, d\sigma(q'),
\end{equation}
where the respective kernels for non-degenerate and M-degenerate Hamiltonians are given by
\begin{equation}
\kernelM{T}{1}={\sump{s, s' \geq 1}{\infty}} \frac{i }{\omega_{s,s'}}\varphi_{s}(q) \varphi_{s'}^{*}(q'),
\end{equation}
\begin{equation}
\kernelM{T}{M}={\sump{s, s' \geq 1}{\infty}}{\sump{r, r' \geq 1}{M}}\frac{i}{\omega_{s,s'}} \varphi_{s,r}(q)\varphi_{s',r'}^{\ast}(q').
\end{equation}
When the energy eigenvalues satisfy equation (\ref{compact}), these kernels are square integrable:
\begin{equation}
\int_{\Omega}\int_{\Omega}\abs{\kernelM{T}{1}}^2\, d\sigma(q)\, d\sigma(q')= {\sump{s, s'\geq 1}{\infty}}\frac{1}{\omega_{s,s'}^2}<\infty,
\end{equation}
\begin{equation}
\int_{\Omega}\int_{\Omega}\abs{\kernelM{T}{M}}^2\, d\sigma(q)\, d\sigma(q')=(M-1)M {\sump{s, s' \geq 1}{\infty}}\frac{1}{\omega_{s,s'}^2}<\infty.
\end{equation}
We know that such operators are compact. The compactness of $\opr{T}_M$, coupled with its self-adjointness, means that $\opr{T}_M$ has discrete spectrum and its eigenvectors span the entire $\mcal{H}$.

\end{document}